# Controlling Magnetic and Electric Nondipole Effects with Synthesized Two Perpendicularly Propagating Laser Fields

Yankun Dou[1], Yiqi Fang[1], Peipei Ge[1] and Yunquan Liu[1,2,3*]

[1] State Key Laboratory for Mesoscopic Physics and Frontiers Science Center for Nano-optoelectronics, School of Physics, Peking University, Beijing 100871, China

[2] Collaborative Innovation Center of Quantum Matter, Beijing 100871, China

[3] Collaborative Innovation Center of Extreme Optics, Shanxi University, Taiyuan, Shanxi 030006, China

**Abstract:** Nondipole effects are ubiquitous and crucial in light-matter interaction. However, they are too weak to be directly observed. In strong-field physics, motion of electrons is mainly confined in transverse plane of light fields, which suppresses the significance of nondipole effects. Here, we present a theoretical study on enhancing and controlling the nondipole effect by using the synthesized two perpendicularly propagating laser fields. We calculate the three-dimensional photoelectron momentum distributions of strong-field tunneling ionization of hydrogen atoms using the classical trajectory Monte Carlo model and show that the nondipole effects are noticeably enhanced in such laser fields due to their remarkable influences on the sub-cycle photoelectron dynamics. In particular, we reveal that the magnitudes of the magnetic and electric components of nondipole effects can be separately controlled by modulating the ellipticity and amplitude of driving laser fields. This novel scenario holds promising applications for future studies with ultrafast structured light fields.

Dipole approximation is widely adopted in intense-light-matter interaction, whose core idea is to ignore the spatial dependence of the driving electric field as well as the effect of the magnetic field component [1,2]. It works successfully in interpreting a larger number of well-known strong-field phenomena, such as above-threshold ionization [3,4], high harmonic generation [5-8] and photoelectron holography [9-11]. With the advances in intense laser technology, the wavelength of laser pulse becomes comparable to the ponderomotive motion of electrons liberated from atoms and molecules [12]. Meanwhile, when the light intensity reaches the regime where the relativistic effect is significant, the magnetic component of the ultra-intense laser pulses in the mid-infrared regime becomes important [13]. In these cases, the dipole approximation becomes invalid and the nondipole effect should be considered [14,15]. Recently, the nondipole effect is experimentally accessible [16] and it has attracted much attention in light-matter interaction, especially in highly nonperturbative and relativistic regimes [17-19]. In photoionization processes, such effects sightly break the forward-backward symmetry of the photoelectron momentum distribution (PMD) [20-23].

The lowest-order nondipole includes the magnetic dipole and the electric quadrupole. For the magnetic dipole effect, it is related to the light pressure effect, i.e., the linear momentum of photons, which drives electrons forward by the Lorentz forces imposed by the magnetic fields [24]. It triggers the research on the sharing of the photon linear momentum between ionized electrons and ions. The underlying mechanism in strong-field ionization has been hotly debated in theory and experiment [25-32]. As for the electric quadrupole effect, it is induced by the spatial dependence of the electric fields [33]. This effect has been discussed in single-photon regime, which results in the forward-backward asymmetry of photoelectron emission [34-36]. In strong-field regime, the electric nondipole effect on strong-field ionization was recently experimentally revealed [37], which can be understood in analogy to the well-known Doppler effect. In most strong-field studies, the driving light field is approximated as a paraxial laser beam [38], and the longitudinal electromagnetic component can be neglected. Therefore, it is usually regarded as a two-dimensional light field, and the laser-induced motion of electrons is confined within the polarization plane of driving fields. The weak longitudinal momentum makes it difficult for electrons to feel the spatial oscillations of the electric field, thus giving rise to the faint nondipole effect that is hard to be experimentally observed. How to enhance and modulate the nondipole effects in strong-field community is a very important issue, but it has so far remained untapped.

In this letter, we propose a robust scheme to enhance and control nondipole effect on strong-field tunneling ionization. As shown in Fig. 1(a), we synthesize two perpendicularly propagating light fields (beam 1 and 2). In this configuration, beam 1 will introduce a strong electric field component with respect to the longitudinal direction of beam 2, and vice versa. It greatly modifies the motion of electrons in both three-dimensional real space and momentum space. In this interaction geometry, driven by three-dimensional light fields, the electron momentum components along the longitudinal directions of two beams will be enhanced. This would result in a strong Doppler effect and action of the magnetic field. Such Doppler effect leads to enhancement of nondipole phenomena. We also study the intrinsic dynamics of the tunneled electrons in the noncollinear synthesized light fields and show how the electric and magnetic nondipole effects alter the electron distributions of strong-field ionization. Moreover, we show that by varying the field parameters of the two beams we can modulate the longitudinal momentum of electrons. Consequently, both the electric and magnetic nondipole effects can be separately controlled.

We start from the Hamiltonian of H atoms in the intense light field [39,40], which expressed as,

$$\mathbf{H} = \frac{1}{2}[\mathbf{p} + \mathbf{A}(\mathbf{r},t)]^2 - V(\mathbf{r}),  \quad (1)$$

where $\mathbf{A}(\mathbf{r}, t)$ is the vector potential of the light fields, and $V(\mathbf{r})$ is the Coulomb potential between the electron and the nucleus. Since the definition of spatial dependent vector potential tightly depends on a certain light propagation direction, here we define the synthesized vector potential for the employed two perpendicularly propagating light fields as

$$\mathbf{A}(\mathbf{r},t) = \mathbf{A}_0(t) + \alpha[(\mathbf{e}_{k1} \cdot \mathbf{r})\mathbf{E}_{01} + (\mathbf{e}_{k2} \cdot \mathbf{r})\mathbf{E}_{02}] + O(\alpha^2), \quad (2)$$

with the unit vectors $\mathbf{e}_{k1}$ and $\mathbf{e}_{k2}$ representing the laser propagation directions of the two beams and $\alpha$ the fine-structure constant. The $\mathbf{e}_{k1} \cdot \mathbf{r}$ and $\mathbf{e}_{k2} \cdot \mathbf{r}$ are the projections of space vector $\mathbf{r}$ in two directions. $\mathbf{A}_0(t)$, $\mathbf{E}_{01}(t)$ and $\mathbf{E}_{02}(t)$ represent the values of the vector potential and the electric field of the two beams at $|\mathbf{r}| = 0$, respectively. The second-order term $O(\alpha^2)$ is ignored. Based on Eq. (2), the electric field and magnetic field can be expressed as $\mathbf{E} = -\partial_t \mathbf{A} - \nabla \phi$, $\mathbf{B} = \nabla \times \mathbf{A}$.

We use the classical trajectory Monte Carlo (CTMC) model to study the nondipole effect of strong-field ionization [41,42]. Here, the ionization probability is given by the adiabatic tunneling Ammosov-Delone-Krainov (ADK) theory [43], in which the initial longitudinal velocity of tunneled electron is assumed to be zero, and the initial transverse momentum, which is perpendicular to the instantaneous electric field at the ionization instant, is described by a rotationally symmetric Gaussian momentum distribution centered at zero momentum. The tunneling exit is given by the effective potential theory [44]. After tunneling, the electrons propagate in the combined external laser field and Coulomb field with the motion governed by Newton's equation $\ddot{\mathbf{r}} = -\mathbf{E} - \mathbf{v} \times \mathbf{B} - \mathbf{r}/r^3$. Here we expand the motion equation to

$$\ddot{\mathbf{r}} = -\mathbf{E}_0(t) - \nabla V(\mathbf{r}) + \mathbf{F}_{n1} + \mathbf{F}_{n2}. \tag{3}$$

The last two terms in Eq. (3) represent a low-order nondipole correction terms of the two beams,

$$\begin{aligned}\mathbf{F}_{n1} &= \alpha\left((\mathbf{e}_{k1}\cdot\mathbf{r})\partial_t\mathbf{E}_{01}(t) - \mathbf{v}\times\{\nabla\times[(\mathbf{e}_{k1}\cdot\mathbf{r})\mathbf{E}_{01}(t)]\}\right) \\ \mathbf{F}_{n2} &= \alpha\left((\mathbf{e}_{k2}\cdot\mathbf{r})\partial_t\mathbf{E}_{02}(t) - \mathbf{v}\times\{\nabla\times[(\mathbf{e}_{k2}\cdot\mathbf{r})\mathbf{E}_{02}(t)]\}\right)\end{aligned} \tag{4}$$

At this point, the dynamic equations can be rearranged as $\ddot{\mathbf{r}} = \ddot{\mathbf{r}}_{E_0} + \ddot{\mathbf{r}}_{E_n} + \ddot{\mathbf{r}}_{B_n} - \mathbf{r}/r^3$, in which $\ddot{\mathbf{r}}_{E_0} = -\mathbf{E}_0(t)$ is the electric dipole term, $\ddot{\mathbf{r}}_{E_n} = -\mathbf{E}_n(\mathbf{r},t) = \alpha\sum_i(\mathbf{e}_{ki}\cdot\mathbf{r})\partial_t\mathbf{E}_{0i}(t)$ is the electric nondipole term, and $\ddot{\mathbf{r}}_{B_n} = -\mathbf{v}\times\mathbf{B}(\mathbf{r},t) = -\alpha\sum_i\mathbf{v}\times\{\nabla\times[(\mathbf{e}_{ki}\cdot\mathbf{r})\mathbf{E}_{0i}(t)]\}$ is the magnetic nondipole term from beam $i$ = 1, 2. As compared to the dipole term, the amplitude of the low-order nondipole term is related to fine structure constant $\alpha$, the reciprocal of the speed of light $\alpha=1/c$, which means that the nondipole signal is several orders of magnitude smaller than the dipole signal [45, 46].

As shown in Fig. 1(a), we calculate the PMDs in two perpendicularly propagating beams with the same angular frequency $\omega$. The light electric fields in the intersection region of these two beams can be written as

$$\begin{aligned}\mathbf{E}_1(\mathbf{r},t) &= \varepsilon_1 E_1 \cos(\omega t + ky)\cdot\mathbf{e}_x + E_1 \sin(\omega t + ky)\cdot\mathbf{e}_z \\ \mathbf{E}_2(\mathbf{r},t) &= E_2 \sin(\omega t + kx)\cdot\mathbf{e}_y + \varepsilon_2 E_2 \cos(\omega t + kx)\cdot\mathbf{e}_z\end{aligned} \tag{5}$$

where $E_i$, $\varepsilon_i$ and $k$ are the amplitude, ellipticity, and wave numbers of the beam $i$ = 1, 2, respectively. Beam 1 is a light field propagating along the $y$-axis, and beam 2 propagates along the $x$-axis. The magnetic field is given by $\mathbf{B}_i = \alpha\mathbf{e}_{ki}\times\mathbf{E}_i$. For the real light fields, the wave numbers $k = \omega/c$ ($\omega$ being the angular frequency, and $c$ being the speed of light in vacuum), while $k$ in the dipole approximation is set as 0. In the simulation, to avoid the additional complexity caused by recollision of the electron with its parent ion, two beams are circularly polarized and have opposite ellipticities, i.e., $\varepsilon_1 = 1$, $\varepsilon_2 = -1$, and $\omega = 0.057$ a.u. (wavelength at 800 nm) and $E_1 = E_2 = 0.06$ a.u. (intensity of laser pulses of ~$2.8\times10^{14}$ W/cm$^2$). Atomic units (a.u.) are used, unless specified. The two beams have the same trapezoidal envelope consisting of a constant plateau of eight cycles and a two-cycle ramp down. One can see that the simulated three-dimensional PMD appears as an oblique elliptical ring shape (blue shape in Fig. 1(a)).

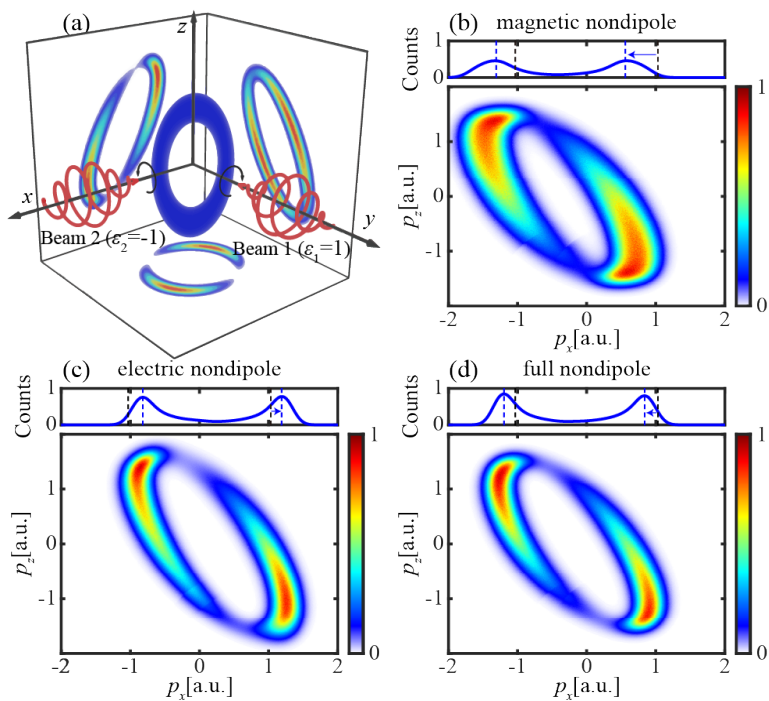

Fig. 1. (a) Illustration of the geometry for controlling the nondipole effect using the two perpendicularly propagating laser pulses. The blue elliptical ring shape indicates the PMDs from strong-field ionization in the synthesized light fields with two opposite ellipticity circularly polarized beams ($\varepsilon_1=1$, $\varepsilon_2=-1$) at 800 nm with the intensity of $2.8\times10^{14}$ W/cm$^2$. The PMDs on the three walls are the projections of the three-dimensional PMD in the dipole approximation. (b)-(d) The PMDs in the $x$-$z$ plane in different cases: only considering the magnetic nondipole effect (b), only considering the electric nondipole effect (c), considering both nondipole effects (d). The corresponding 1D distributions of the momentum along $p_x$ direction are also plotted. The dashed lines represent the change of the peak position of the 1D distribution before (black) and after (blue) considering the nondipole effect.

To comprehensively understand the nondipole effects on strong-field ionization, we have disentangled the nondipole effects and simulated the following three cases: (i) only considering the magnetic nondipole effect, (ii) only considering the electric nondipole effect, (iii) considering both the magnetic and electric nondipole effects. Here, to clearly visualize the nondipole effect, we focus on the PMDs in the *x-z* plane for the above three cases, as shown in Figs. 1(b)−(d). The corresponding 1D distributions of the momentum along $p_x$ direction are presented. The dashed lines represent the change of the peak position of the 1D distribution before (black) and after (blue) considering the nondipole effect. It is worth noting that the nondipole effect modifies both the *x*- and *y*-direction momentums of the liberated electrons. These two components are equivalent. Without loss of generality, we therefore concentrate on the influence of nondipole effect on the photoelectron momentum along the *x*-direction, or, *x-z* plane. Moreover, since the synthesized light fields impede the recollision, the influence of the Coulomb field on the results is trivial.

As seen in Fig. 1(d), when full nondipole effects are included, the PMD shifts along -$p_x$ direction. The average offset is approximately -0.195 a.u., which is significant compared to previous nondipole phenomena. In order to figure out the nondipole effects, the individual magnetic or electric nondipole effect is considered lonely. We find that the offset from the magnetic nondipole effect is much larger than that from the electric nondipole effect and they are in opposite directions, as shown in Figs. 1(b) and 1(c). Their values are -0.379 a.u. and 0.183 a.u.. Thus, the momentum offset including full nondipole effects are manifested as the superposition of the two nondipole effects.

To interpret this interesting result, we have analyzed the sub-cycle dynamics of tunneled electrons by tracing the corresponding electron trajectories with the dipole and nondipole effects. As shown in Fig. 2(a), the electron trajectories in different cases show different deviations in the *x* direction as compared to the trajectory in the dipole approximation.

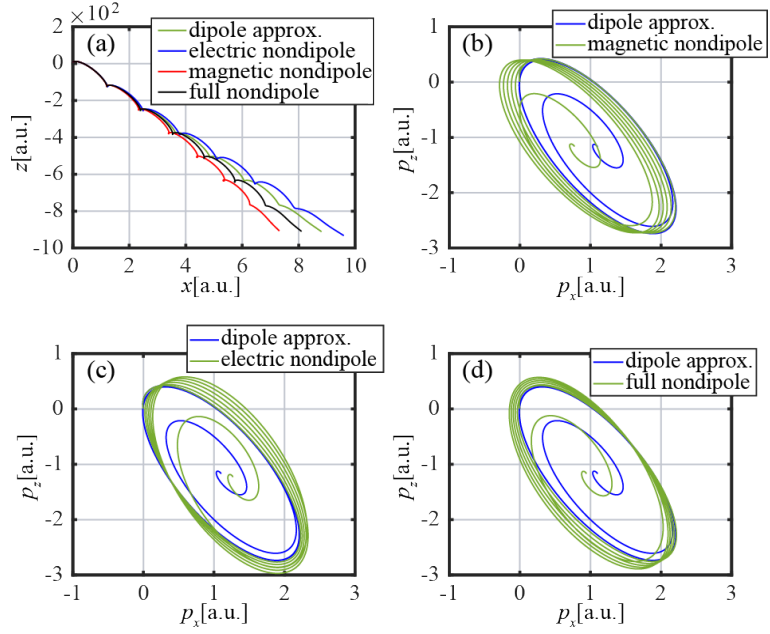

Fig. 2. (a) Tunneled electron trajectories projected onto the *x-z* plane in different cases. (b)-(d) The comparison of the electron momentum evolution projected on the *x-z* plane among the dipole approximation (blue line) and different nondipole cases (green line). Note that the polarization plane is tilted in three-dimensional momentum space.

Under the dipole approximation, the electron momentum evolution is confined within a plane which is parallel to the polarization plane of the synthesized light fields. The longitudinal momentum of the electron keeps unchanged during the momentum evolution. Note that the polarization plane is tilted here. As shown in Figs. 2(b) and 2(c), when considering the individual magnetic or electric nondipole effect, the electrons move out of this plane, following a spiral trajectory. Moreover, the derivation of the electron momentum is continuously accumulated during the propagation. This causes a significant shift in the final electron longitudinal momentum at the end of pulse. The effect from the magnetic nondipole is stronger, resulting in a larger final electron momentum offset. Under the simultaneous action of the two nondipole effects, the offset of the momentum evolution is manifested as a superposition of two nondipole effects and the influence of magnetic nondipole dominates. It is consistent with the phenomena shown in Figs. 1(b)−(d).

One can think that there is a constant force pointing out of the polarization plane, and the directions of such constant forces induced by the magnetic and electric nondipole effects are opposite. To inspect the origin of such constant forces, we then perform a qualitative analysis by solving the dynamic equations of electrons in the synthesized light fields. We substitute the electric field expressions in Eq. (5) into Eq. (3), and obtain the nondipole-corrected dynamic equations of electrons in the synthesized light fields, with the nondipole terms written as

$$\ddot{\mathbf{r}}_{E_n} + \ddot{\mathbf{r}}_{B_n} = \begin{cases} \left(C_E[\cos(2\omega t)-1] + C_B[\cos(2\omega t)+1] + O_2\right) \cdot \mathbf{e}_x \\ \left(C_E[\cos(2\omega t)+1] + C_B[\cos(2\omega t)-1] + O_1\right) \cdot \mathbf{e}_y \\ \alpha(E_1 E_2/\omega)(1-\varepsilon_1\varepsilon_2)\sin(2\omega t) \cdot \mathbf{e}_z \end{cases} \quad (6)$$

The constant terms $C_E = (\alpha/2)\varepsilon_1(E_1 E_2/\omega)$ and $C_B = (\alpha/2)(E_1 E_2/\omega)(\varepsilon_1-\varepsilon_2)$ are from the electric and magnetic nondipole effects, respectively, and $O_i = (\alpha/2)(E_i^2/\omega)(\varepsilon_i^2-1)\sin(2\omega t)$ is the oscillating term from magnetic nondipole effect. The oscillating part $C_E \cdot \cos(2\omega t) + C_B \cdot \cos(2\omega t) + O_i$ has a little contribution to the momentum shift due to the fine structure constant $\alpha$. However, the constant parts $C_E$ and $C_B$ can cause the shift of momentum to accumulate along $p_x$ and $p_y$ direction during the electron evolution as mentioned earlier. When the two beams are circularly polarized and the ellipticities of them are opposite, i.e., $\varepsilon_1 = 1$ and $\varepsilon_2 = -1$, the constant terms satisfy $C_B = -2C_E = -2(\alpha/2)(E_1 E_2/\omega)$. Thus, the offset of momentum distribution caused by the case i is twice as large as that caused by the case ii, and they are in opposite directions. As for the case iii, the two constant terms are added, causing the offset to change, as described earlier. The qualitative analysis via the solution of the dynamic equation agrees well with the simulation. Especially, for the strong-field ionization driven by single light beam, the amplitude of beam 2 propagating in the $x$-direction satisfies $E_2 = 0$. Then, the constant terms $C_B$ and $C_E$ that cause the PMD to shift continuously are zero. The nondipole effect of the light fields on electrons only originates from the oscillation term $O_1$ along the $y$ direction. This indicates that the previously observed forward momentum shift is caused by the magnetic nondipole effect (or photon's linear momentum) of beam 1, and its weak features are also intuitive.

From $C_E$ and $C_B$, we find that the primary factors influencing the nondipole phenomena are the ellipticity $\varepsilon_i$, amplitude $E_i$ and frequency $\omega$ of two beams. This means that one can control the magnitude of the nondipole effects by modulating these laser parameters. By inspecting the two constant terms, one can find that the constant term from the electric nondipole effect $C_E = (\alpha/2)\varepsilon_1(E_1 E_2/\omega)$ is only related to the ellipticity of beam 1, but $C_B = (\alpha/2)(E_1 E_2/\omega)(\varepsilon_1-\varepsilon_2)$ is affected by the ellipticities of both beams.

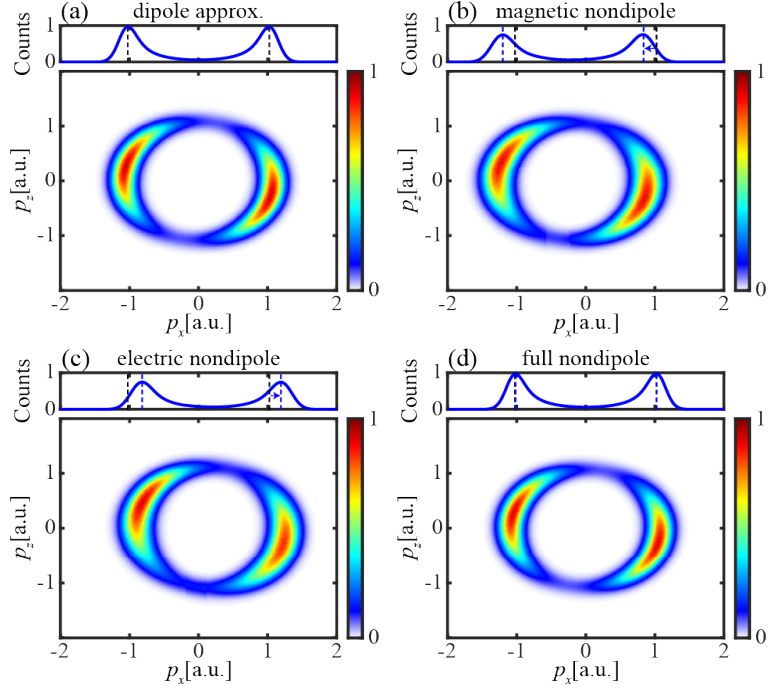

Fig. 3. When beam 2 is horizontally polarized ($\varepsilon_2=0$), the PMDs in the *x-z* plane are shown for different cases: under the dipole approximation (a), only considering the magnetic nondipole effect (b), only considering the electric nondipole effect (c), and considering both two nondipole effects (d). On the top of each panel, the corresponding 1D distribution of the momentum along $p_x$ direction is displayed. The dashed lines represent the peak position of the 1D distribution before (black) and after (blue) considering the nondipole effect.

Here, we study a special case of $E_1 = E_2 = 0.06$, $\varepsilon_1 = 1$, and $\varepsilon_2 = 0$ (beam 2 is horizontally polarized). In this case, the two constant terms satisfy $C_B = -C_E$. The three-dimensional PMD manifests a donut-like shape that slopes along the *y*-axis. As seen in Figs. 3 (b) and (c), the individual magnetic or electric nondipole effect induces PMD to shift into the opposite direction, and the momentum offsets are similar. This leads to a consequence that when both nondipole effects are considered, the momentum offset vanishes and the PMD follows almost the same as that in dipole approximation, as shown in Figs. 3 (a) and (d). It also implies that the two constants $C_B$ and $C_E$ are practical in characterizing the nondipole effects in arbitrary three-dimensional synthesized light fields. Here, we show that the electric and magnetic nondipole effects are comparable, which breaks the impression that electric nondipole effect is much smaller than magnetic nondipole effects.

To disentangle the offset contributions from the nondipole effects, we define the mean momentum $\langle p_x \rangle$, which is obtained by taking the ionization probability of 1D momentum distribution into account for each momentum $p_x$. This allows us to quantify the magnitude of the nondipole effects. When controlling the field ellipticity of beam 2 ($\varepsilon_2$), one can clearly visualize the mean momentum $\langle p_x \rangle$ modulation with respect to the nondipole effects, as shown in Fig. 4(a). As $\varepsilon_2$ changes from -1 to 1, $\langle p_x \rangle$ keeps at zero in the dipole approximation. When including the electric nondipole effect, $\langle p_x \rangle$ keeps around 0.17 for different ellipticities. However, for the case including the magnetic nondipole effect, the mean $\langle p_x \rangle$ monotonically increases from -0.38 to 0. Obviously, the magnetic nondipole effect can be sensitively controlled by monitoring the field helicity.

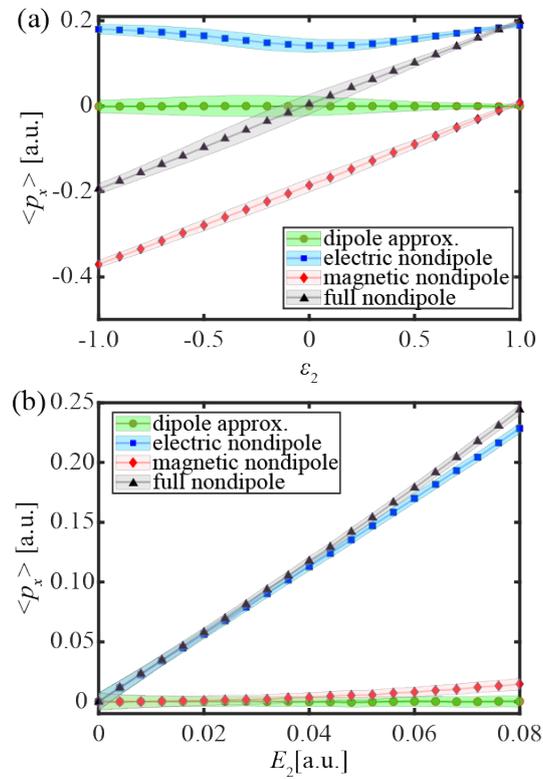

Fig. 4. The calculated mean momentum $\langle p_x \rangle$ of 1D distribution of the momentum along the $p_x$ direction, which can be used to characterize the offset contributions from the nondipole effects and to quantify the magnitude of the nondipole effects. (a) The mean momentum $\langle p_x \rangle$ with respect to the ellipticity of beam 2 in dipole approximation (green), only with the magnetic nondipole (red), only with the electric nondipole (blue), and with both nondipole terms (black) is displayed. (b) The mean momentum $\langle p_x \rangle$ with respect to the amplitude of beam 2 in the four cases mentioned above.

For the full nondipole effect, it can be expressed as the superposition of two nondipole effects, $C_F = C_E - C_B = (\alpha/2)\varepsilon_2(E_1 E_2/\omega)$. This corresponds to the fact that $\langle p_x \rangle$ increases from negative to positive when increasing $\varepsilon_2$ from $-1$ to $1$. In particular, in the case of $\varepsilon_2 = -1$, the influence of the magnetic nondipole effect dominates over the electric nondipole effect. Therefore, the forward-backward asymmetry is mainly caused by the magnetic nondipole effect. In the case of $\varepsilon_2 = 0$, one can observe that the $\langle p_x \rangle$ experiences opposite modulations by the magnetic and electric nondipole effects. The two effects are canceled out when simultaneously considering contributions from the two nondipole terms, causing a forward-backward symmetry on the $\langle p_x \rangle$. The ellipticity-dependent results show that one can modulate the magnetic nondipole effect individually by changing the ellipticity of beam 2.

Moreover, the amplitude of the laser pulses $E_i$ is also a key factor of controlling the nondipole effects. Then, we study the field amplitude of beam 2 on the nondipole effects. Here, both beams are still right-rotating circularly polarized (i.e., $\varepsilon_1=\varepsilon_2=1$). In this case, the magnetic nondipole effect is almost suppressed as shown in Fig. 4(a). The total nondipole effect mainly originates from the electrical nondipole effect. Likewise, we calculate the mean momentum $\langle p_x \rangle$ with respect to the amplitude $E_2$, as shown in Fig. 4 (b). Since the magnetic nondipole effect is faint, its corresponding curve overlaps with the result of the dipole approximation. In addition, when the light intensity is high, the oscillation terms in Eq. (5) also cause a minor contribution to the PMDs shift. With the increase of laser amplitude $E_2$, one can see that $\langle p_x \rangle$ affected by the electric nondipole effect increases monotonically and gradually deviates from the results in dipole approximation. Therefore, by changing the field amplitude of the noncollinear three-dimensional synthesized light fields, one can control the electric nondipole effect individually and further control the total nondipole effect.

In conclusion, we have studied the enhancement and controlling of the nondipole effect in strong-field ionization by using the synthesized two perpendicularly propagating beams. It is revealed that the nondipole effects give rise to significant shifts of PMDs in strong-field ionization. The nondipole-corrected dynamic equations of electrons in the noncollinear synthesized light fields are derived. Then, we show that the electric and magnetic nondipole effects can be separately controlled and enhanced by tuning the amplitude and ellipticity of the laser field. Note that the electric nondipole effect is typically one order of magnitude weaker than the magnetic nondipole effect

[47]. Astonishingly, the electric nondipole effect is significantly enhanced here. It could be comparable to or even stronger than the magnetic nondipole effect. It is worth stressing that the spatial phase fluctuation between two laser pulses have more or less influence on the experimental results. This effect may be avoided as far as possible by reducing the focal volume or by choosing appropriate laser conditions and observables. We hope to implement this work experimentally in the near future. Such controlling scenario should attract much attention in ultrafast and strong-field physics, and this would have a promising perspective on controlling nondipole effects for more molecules with bi-circular fields [48,49].

*Acknowledgments*. This work was supported by the Key R&D Program of China (Grant No. 2022YFA1604301), and the National Natural Science Foundation of China (Grant No. 92050201, 92250306, and 12204018).